\documentclass[pre,twocolumn,aps,superscriptaddress,showpacs,loatfix]{revtex4}

\usepackage{graphicx}
\usepackage{amsmath}
\usepackage{amsfonts}
\begin{document}

\title{Transport driven by biharmonic forces:  impact of  correlated thermal noise}

\author{M. Machura}
\affiliation{Institute of Physics, University of Silesia,
40-007 Katowice, Poland}

\author{J. \L uczka}
\affiliation{Institute of Physics,  University of Silesia,
40-007 Katowice, Poland}

\begin{abstract}
 We study   an inertial Brownian particle  moving in  a symmetric 
periodic substrate,  driven by a zero-mean biharmonic force and correlated thermal noise.  
 The Brownian motion  is described in terms of  a Generalized
Langevin Equation with an exponentially correlated Gaussian  noise term, 
 obeying the fluctuation-dissipation theorem. We analyse impact of non-zero correlation time of thermal noise  on transport properties of the Brownian particle.   We  identify  regimes where the increase of the correlation time intensifies long-time  transport  of the Brownian particle. 
The opposite effect is also found: longer correlation time reduces the stationary velocity of the particle.  The correlation time induced multiple current reversal  is  detected. We reveal that thermal noise of non-zero correlation time can radically enhance  long-time velocity  of the Brownian particle in regimes where in the white noise limit the velocity  is extremely small.  
All transport properties can be tested in the setup consisting of a resistively and
  capacitively shunted Josephson junction device.
\end{abstract}


\maketitle

\section{Introduction} 

Directed transport of particles  can be driven by zero-average forces.  The second principle of thermodynamics excludes it for systems at thermodynamic  equilibrium. Therefore  the system has to be at  a nonequilibrium state and, moreover,  some  symmetries of  the system have to be  broken as   
e. g. the reflection symmetry  of the spatially periodic potential  or time-symmetry of the external force. 
Presence of both nonequilibrium conditions and asymmetry  usually  leads to transport and these two 
constituents  form the ratchet  concept.  Transport in ratchet systems can be powered by mechanical, electrical, optical,  chemical  or electronic means. 
A large variety of models have been proposed and realized experimentally as solid state devices, cold atoms in optical lattices, superconducting devices, geometrically asymmetric lattices, to mention but a few, see the review \cite{RMP2009}.  
                                                          
The zero-average force  can be deterministic or stochastic.  An example of  such a deterministic force is  a time-periodic AC driving consisting of one or several harmonics.  In some cases there is possibility of generating a charge or particle DC  current from pure AC-driving, without the presence of an explicit static bias.  
  The well-known phenomenon is harmonic mixing \cite{schneider}.  A paradigmatic model of this phenomenon is based on  the Newton-Langevin equation which describes a particle moving in a spatially periodic potential (which mimics e.g.  the motion of a charge   particle in a crystal) and driven by biharmonic force $F(t)$ of two components of frequencies $\omega_1$ and $\omega_2$  and amplitudes 
$F_1$ and $F_2$, respectively,  
\begin{equation}
\label{F}
F(t) =  F_1  \sin(\omega_1 t)  + F_2 \sin(\omega_2 t + \phi), 
\end{equation}
where $\phi$ is the phase-lag of two signals.  Transport in such systems has been studied 
in various contexts,   
mainly in the overdamped regime \cite{Borromeo2005a}, for moderate damping \cite{Breymayer1984}, 
also in  dissipative quantum systems \cite{igor}, 
both experimentally and theoretically for cold atoms in the optical lattices
 \cite{Renzoni2005,Renzoni2008,Denisov2010},
and for  driven Josephson junctions \cite{Monaco1990}.

The sketch  of the paper is as follows.  In Sec. II  we briefly present main elements of modelling for typical systems exhibiting the ratchet effects  when thermal fluctuations are  described by Gaussian white noise.       Sec. III contains the formulation of the corresponding model of non-Markovian dynamics for systems  driven by biharmonic signals and correlated thermal  noise.   In Sec. IV we analyse in detail influence of correlation time  of thermal  noise on transport  properties.    
  Finally,  Sec. V  provides summary.

\section{Paradigmatic  Model} 

An  archetype of a ratchet  is based on the Newton-Langevin equation 
%
\begin{equation}
\label{LE}
m \ddot{x} + \gamma \dot{x}  = - V'(x) + G(t) + \xi(t), 
\end{equation}
where  $ m $ is a mass of the  particle,  $ \gamma $ is the 
friction coefficient and  $ V(x) $ is a spatially periodic potential of period $L$, i.e. $V(x)=V(x+L)$.  The deterministic force $G(t)$  is a non-biased  function of zero average over some time-interval. The last term $\xi(t)$ is a stochastic force which can model thermal equilibrium fluctuations and/or 
non-equilibrium  noise. 

Variations of this equation are  unlimited. Putting  formally $m=0$, the overdamped system is modelled. 
In the case $m\ne 0$, inertial effects can be investigated.  The potential $V(x)$ can be symmetric or asymmetric, the force  $F(t)$ can be symmetric or not, the same for $\xi(t)$ unless it mimics thermal noise, which is always symmetric.  Problem of symmetry of the  stochastic driving $\xi(t)$ is discussed in Ref. \cite{euro2}. 

 If the process $\xi(t)$ describes thermal equilibrium  fluctuations then  $\xi(t)$ is  zero-mean, Gaussian
white noise  with the Dirac delta auto-correlation function
\begin{equation}
\label{white}
\langle \xi(t)\xi(s)\rangle = 2 \gamma k_B T_0 \, \delta(t-s), 
\end{equation}
where $k_B$ the Boltzmann constant and  $T_0$ denotes the temperature.  Note that this process in not correlated  and in consequence its correlation time $\tau_c =0$.  The noise intensity  factor 
$ 2\gamma k_B T_0$  follows from the fluctuation-dissipation theorem  \cite{kubo66}. 
Gaussian white noise generates mathematically tractable models. 
In many real  systems thermal noise is approximately white, meaning that the power spectral density is nearly equal throughout the frequency spectrum.   Additionally, the amplitude of the signal has very nearly a Gaussian probability density function. However, 
an infinite-bandwidth white noise signal is purely a theoretical construction. By having power at all frequencies, the total power of such a signal is infinite and therefore impossible to generate. In practice, however, a signal can be "white" with a flat spectrum over a defined frequency band and it 
is a good approximation of many real-world situations.  E.g. in electrical systems, 
the white noise  modelling is correct  at any practical radio frequency in use (i.e. frequencies below about 80 GHz). In the most general case, which includes up to optical frequencies, the power spectral density of the voltage across the resistor  depends on frequency and the white noise  approximation fails.

In the paper, we study transport in a spatially symmetric substrate and the corresponding  potential is reflection symmetric, i.e. $V(x_0 +x) = V(x_0 -x)$  for some fixed value $x_0$.  One of the simplest form of such a potential is given by the sinusoidal form
\begin{equation}
\label{pot}
V(x) =  \Delta V \sin (2\pi x/L) 
\end{equation}
of the period $ L $ and the barrier height $ 2 \Delta V $.
The non-equilibrium driving $ G(t) $  is a biharmonic force and  is chosen  as a particular case of Eq. (\ref{F}), namely,   
\begin{equation}
\label{gt}
G(t) =  A [ \sin(\Omega t)  + \varepsilon \sin(2 \Omega t + \phi) ].
\end{equation}
Here $ A $ is the  amplitude of the first harmonic,  the factor $ \epsilon $ scales the second harmonics, so 
that it has the resulting  amplitude  $   \varepsilon A $.  The angular 
frequency  $ \Omega $  determines the time period $ T = 2 \pi / \Omega $ of the driving and  
$ \phi $  controls the phase shift between  two components of  the biharmonic signal (\ref{gt}), 
see Fig. \ref{figpot} for details.

The Langevin equation {(\ref{LE})   with the potential (\ref{pot}) has a similar form  
as an equation describing dynamics of  the phase
difference $\Psi=\Psi(t)$  between the macroscopic wave functions of the
Cooper pairs on both sides of the  resistively and capacitively
shunted Josephson junction, which is well known in the literature as  the Stewart-McCumber
model \cite{stewart,junction},  namely, 
\begin{eqnarray} \label{JJ1}
\Big( \frac{\hbar}{2e} \Big)^2 C\:\ddot{\Psi} + \Big( \frac{\hbar}{2e} \Big)^2 \frac{1}{R} \dot{\Psi}
+ \frac{\hbar}{2e} I_0 \sin \Psi \nonumber\\ = \frac{\hbar}{2e} I(t) + \frac{\hbar}{2e}
\sqrt{\frac{2 k_B T}{R}} \:\xi (t) .
\end{eqnarray}
The left hand side is the total current through the junction. It is a sum of  three additive
current contributions: a displacement current accompanied with a  capacitance $C$ of the junction, a normal (Ohmic) current characterized
by the normal state resistance $R$ and a Cooper pair tunnel supercurrent characterized by
the critical current $I_0$.   In the right hand side, $I(t)$ is an external current. 
The Johnson noise $\xi(t)$ plays a role of thermal equilibrium noise  associated with the resistance  $R$ 
and is modelled by Gaussian white noise of zero mean and the intensity  taken according to the fluctuation-dissipation theorem.  
It is an evident correspondence between two models:  the coordinate $x=\Psi -\pi/2$,    the mass $m=(\hbar /2e)^2 C$,   the friction coefficient $\gamma =
(\hbar/2 e)^2(1/R)$, the barrier height is determined by the relation  $\Delta V = (\hbar/2 e) I_0$ and the
period $L=2\pi$.  
 The biharmonic signal $G(t)$ in Eq. (\ref{gt}) corresponds to the external  current $I(t)$ . 
The velocity $v=\dot x$ corresponds to  the voltage $V$ across the junction. 
This correspondence allows for testing all transport properties of the system  (\ref{LE}) and its generalizations by  the setup consisting of a resistively and
  capacitively shunted Josephson junction device.

\begin{figure}[htbp]
\includegraphics[width=0.9\linewidth]{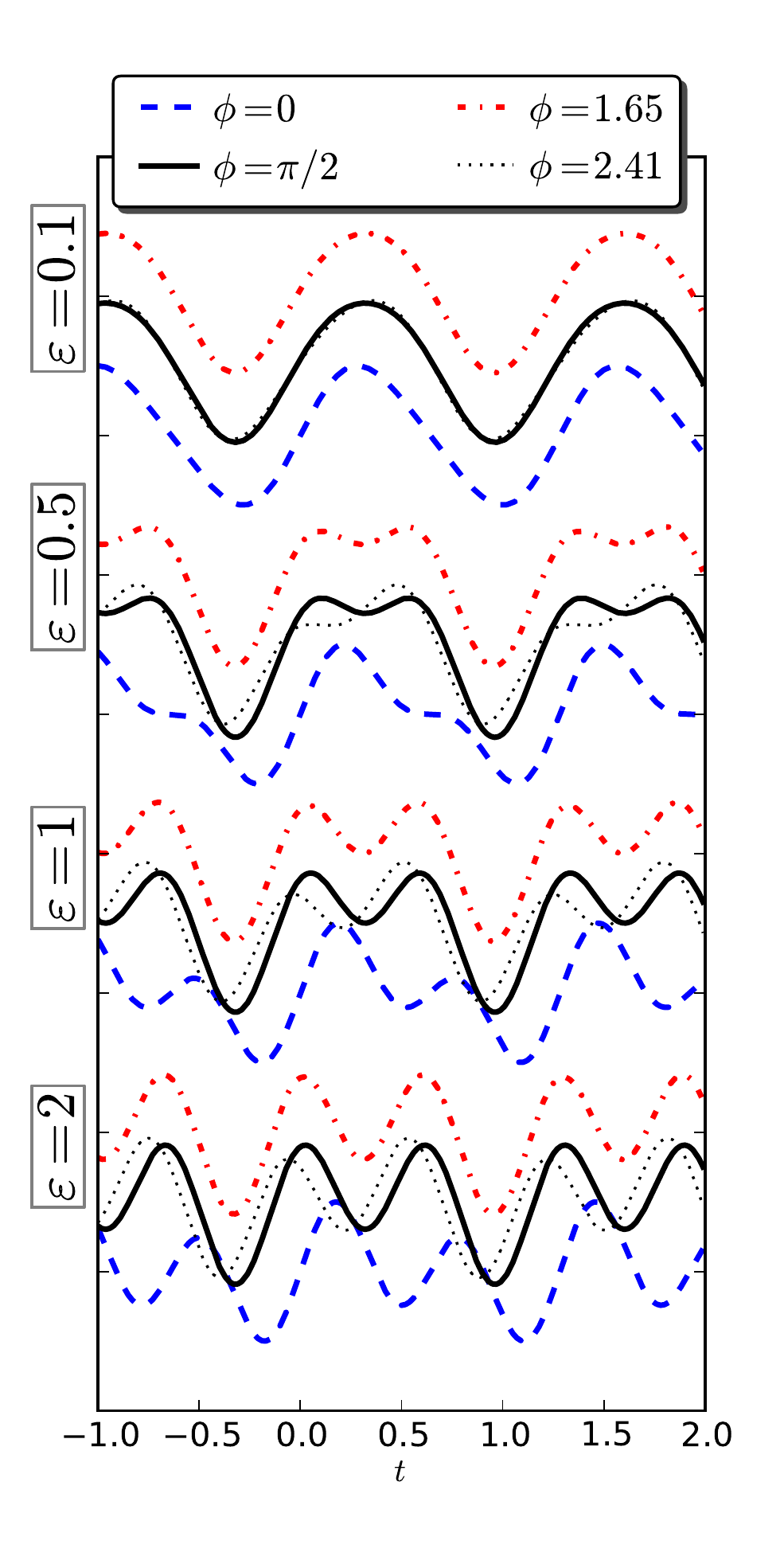} 
\caption{(color online) Biharmonic external ac driving
$G(t) = A[\sin(\Omega t) + \varepsilon \sin (2\Omega t + \phi)]$
plotted for amplitude $ A = 1 $ and frequency $\Omega=1$,
for four different second harmonic  amplitude 
$\varepsilon=0.1, 0.5, 1, 2$ and four phase shifts 
$ \phi =  0, \pi /2, 1.65, 2.41$ which corresponds to the
later commented figures (see \ref{fig3} and \ref{fig4} for
details).}
\label{figpot}
\end{figure}

\section{Non-Markovian dynamics }

If  the correlation time  $\tau_c$  of thermal noise
is  much smaller than the smallest characteristic (deterministic) time $\tau_x$  in  the considered  system then the white noise modelling is a good approximation. 
However, there are  systems where this relation is not satisfied: the noise correlation time can be of  order or  greater than $\tau_x$  \cite{jungACP,chaos,goychuk2005}. In this case
we shell not assume the Markovian dynamics but rather invoke the Generalized 
Langevin Equation (GLE)  for the time evolution of the system (\ref{LE}).

When the correlations comes into play the dynamics becomes non-Markovian and exhibits the
memory-friction described by the GLE in the form 
\cite{kubo66,zwan,chaos,LNP}
\begin{eqnarray}
\label{Gen}
m\ddot x(t) + \int_0^t K(t-s) \dot x(s)\;ds  = - U'(x(t), t) + \xi(t),
\end{eqnarray}
where $\xi(t)$ is a  stationary, zero-mean Gaussian stochastic process modelling correlated thermal noise and the full potential takes the form
\begin{eqnarray}
\label{fullpot}
U(x, t) = V(x) - A [\sin(\Omega t) + \varepsilon \sin(2 \Omega t +\phi) ] x.
\end{eqnarray}
The auto-correlation function $C(t-s)$ of the correlated Gaussian thermal noise $\xi(t)$
is related to the memory kernel $K(t)$ via the fluctuation-dissipation relation 
\cite{kubo66,zwan,chaos,LNP}:
\begin{eqnarray}
\label{mom}
C(t-s) = \langle \xi(t) \xi(s) \rangle = k_B T_0 \, K(|t-s|) . 
\end{eqnarray}
Let us note that for more realistic modelling based on correlated thermal fluctuations we have to pay the price: Eq. (\ref{Gen}) is an integro-differential equation with a Gaussian random term which is difficult to handle by analytical or numerical means. The resulting process is neither Gaussian nor Markovian. 
In a general case, we do not know   even  a master equation for the single-event non-Markovian
 probability $p(x, \dot x,t)$ of the process determined by Eq.  (\ref{Gen}).  However, the problem is much easier to treat if the non-Markovian process $x(t)$ is a projection of a higher-dimensional  Markovian process.  In other words, it can be embedded in a higher (but finite) dimensional phase space  and then  Eq. (\ref{Gen}) can be converted to a set of first-order differential equations of Langevin type.  It can be done if the kernel  $K(t)$ obeys an ordinary differential equation with constant coefficients.

\subsection{Correlated thermal noise }

 Gaussian thermal noise $\xi(t)$ is  completely defined by its 
 correlation function $C(t)$.   White noise corresponds to  (\ref{white}). 
 Examples of correlated noise are: 
the exponentially correlated noise  \cite{ornst},  harmonic noise \cite{lutz},  algebraically correlated noise \cite{srokowski,goj}.   Other variations of noise  are also considered, as e.g.  when $C(t)$ is 
a sum of  exponentials \cite{goj,kupfer,bao}. 

The simplest correlated noise is exponentially correlated noise (because of the smallest number of parameters) 
known as the Ornstein-Uhlenbeck (O-U) stationary stochastic Markov process for $\xi(t)$ \cite{chaos}.  Its  correlation function  reads
\begin{eqnarray}  \label{o-u}
\langle \xi(t) \xi(s)\rangle = k_BT_0 K(|t-s|)
=\frac{\gamma k_BT_0}{\tau_c} \;e^{-|t-s|/\tau_c}.
\end{eqnarray}
Here $ \tau_c $ is the correlation time of the O-U process.  When the correlation time 
$ \tau_c \to 0 $  the correlation function (\ref{o-u}) tends to the correlation function (\ref{white}) and the O-U process tends to white noise.  Then  Eq. (\ref{Gen}) reduces to  Eq. (\ref{LE}). Below, we study the 
case of  O-U noise.

\subsection{Markovian embedding dynamics}

For the case of the O-U noise, the integral kernel (\ref{mom}) is an exponential function of the time.  The exponential function obeys a first-order differential equation with a constant coefficient. Therefore  we can convert the integro-differential equation (\ref{Gen}) into a set of ordinary stochastic differential equations. To this aim  we  define the auxiliary  stochastic process
\begin{eqnarray}
\label{y(t)}
w(t) = \frac{\gamma}{\tau_c} \int_0^t \mbox{e}^{-(t-s)/\tau_c}\dot x(s)\;ds,
\end{eqnarray}
which is an integral part of Eq. (\ref{Gen}). 
By means of the above relation the GLE (\ref{Gen}) can be rewritten in the form 
\begin{eqnarray}
\label{4eq}
m \dot v(t)&=& - U'(x(t), t) -w(t) + \xi (t),   \\
\dot x(t)&=&v(t),  \\
\dot w(t)&=& -\frac{1}{\tau_c} w(t) +\frac{\gamma}{\tau_c} v(t),  \\
\label{O-U}
\dot \xi (t)&=& -\frac{1}{\tau_c} \xi (t)
+\frac{1}{\tau_c} \sqrt{2\gamma k_BT_0} \; \Gamma(t),
\end{eqnarray}
The last equation (\ref{O-U}) corresponds to the O-U process with the
exponential correlation function  in Eq. (\ref{o-u}) \cite{pha}. The  stationary stochastic process 
$ \Gamma (t) $ describes Gaussian white noise  of zero mean and 
correlation function $ \langle \Gamma(t) \Gamma(s) \rangle = \delta(t-s) $. 
 So, we embedded a non-Markovian process in a 4-dimensional  space in which   the process is Markovian. It is not a minimal dimension. It can be further reduced to the 3-dimensional space. To do it. let us note that  for  the linear combination $z(t) = \xi(t)-w(t)$ we are able to 
subtract the two last relations and reproduce the three coupled 
Langevin equations \cite{KosLucHan2009}, namely, 
\begin{eqnarray}
\label{x}
\dot x(t) &=& v(t),\\
\label{v}
 \dot v(t) &=&  -\frac{1}{m} U'(x(t), t) +\frac{1}{m} z(t), \\
 \label{z}
\dot z(t) &=& -\frac{1}{\tau_c}z(t) - \frac{\gamma}{\tau_c} v(t)
+\frac{1}{\tau_c} \sqrt{2\gamma k_BT_0}\; \Gamma(t).
\end{eqnarray}
In the following we shall  analyse this set of  
three coupled equations. 


\subsection{Dimensionless variables} 
The natural length scale for the system is settled by the period $L$ of the potential 
$V(x)$. For the adequate time scale we have to consider several time scales
\cite{bio}. Here we define the characteristic time as
$ \tau_0^2 = mL^2/\Delta V$ which is convenient for studying inertial effects.  It  can be obtained  from Eq. (\ref{LE}) by comparing the inertial term  (with mass $m$) to  the potential force  $V'(x)$  and  inserting in both sides  the characteristic quantities,  for detail see Ref.  \cite{bio}.  
It  leads to the scaling used throughout this paper
\begin{eqnarray}
\label{scaling}
X= \frac{x}{L}, \qquad \hat{t} = \frac{t} {\tau_0},
\end{eqnarray}
and finally to the rescaled evolution equations
\begin{eqnarray}
\label{X}
\dot X &=& Y,    \\
\label{Y}
\dot Y &=& -W'(X) + a [\sin(\omega \hat{t}) + \varepsilon \sin(2 \omega \hat{t}+ \phi)] + Z, \;\;\; \\
\dot Z &=& -\frac{1}{\hat{\tau_c}} Z - \frac{\hat{\gamma}}{\hat{\tau_c}} Y
+\frac{1}{\hat{\tau_c}} \sqrt{2\hat{\gamma} D}\;  \hat{\xi}(\hat{t}),
\label{Z}
\end{eqnarray}
where the explicit form of remaining coordinates reads
\begin{eqnarray}
\label{YZ}
Y = \frac{\tau_0 }{L}\; v, \qquad Z  = \frac{L} {\Delta V} \; z.
\end{eqnarray}
The dot and prime denote the differentiation with respect to the scaled time
$ \hat{t} $ and the argument respectively. Rescaled velocity is designated by
$ Y({\hat t}) $ and $ Z({\hat t}) $ symbolizes the dimensionless random force.
The remaining re-scaled parameters are:
\begin{enumerate}
\item the friction coefficient $\hat{\gamma} = (\gamma / m) \tau_0 =
\tau_0 / \tau_L$ is defined by the ratio the two characteristic times of the
GLE - previously defined $\tau_0$ and the relaxation time of the velocity degree
of freedom $\tau_L = m/\gamma$,
\item the correlation time ${\hat{\tau_c}}= \tau_c/\tau_0$,
\item the potential $W(X)=V(x)/\Delta V = W(X+1) = \sin (2\pi X)$ possesses 
the unit period and barrier height $\Delta \hat{V}=2$,
\item the amplitudes of the external deterministic stimulus scales as 
$a = L A / \Delta V$ and the 
frequency $\omega = \Omega \tau_0$ (or the equivalent period $T=2\pi/\omega$),
\item the zero-mean white noise $\hat{\xi}(\hat{t})$ is correlated as
$\langle\hat{\xi}(\hat{t})\hat{\xi}(\hat{s})\rangle=\delta(\hat{t}-\hat{s})$
with a re-scaled noise intensity $D = k_B T_0 / \Delta V$ and can be interpreted 
as the ratio of two energies, thermal energy and  half of barrier height of the 
potential $V(x)$.
\end{enumerate}

From now on we will use only above dimensionless variables and
shall omit the ``hats'' in all quantities of  Eqs.  (\ref{X})-(\ref{Z}).

\subsection{Method of analysis} 

The analytical methods to handle nonlinear Brownian equations with memory friction 
are unknown to our knowledge. Therefore we will explore the peculiarities of the 
system (\ref{X})-(\ref{Z}) by means of numerically calculated long-time transport 
characteristics. In particular we shall focus on the current defined by the
long-time averaged velocity $ v \equiv \langle Y \rangle$.
The averaging is performed in  the following way: First the numerical mean over 
at least $ 10^3 $ realizations of the GLE is calculated. This yields a time 
dependent quantity which we next average over one period of the external field
$ T $. In the process of calculating averages we must make sure that we initiate 
the process evolutions with unbiased initial conditions since the simulated 
asymptotic, long time dynamics is not necessarily ergodic.
We choose all initial positions and velocities to be uniformly distributed 
over one potential period  $[0, 1]$ and the interval $v \in [-2,2]$,  
respectively. From the technical point of view we have employed Stochastic 
Runge--Kutta algorithm of the $ 2^{nd} $ order \cite{PhysRevA.45.600}  
with the time step of $[10^{-3} \div 10^{-4}]T$. All numerical calculations have 
been performed using CUDA environment on desktop computing processor NVIDIA 
Tesla C1060 \cite{cuda}.


%
\begin{figure}[htbp]
\includegraphics[width=0.90\linewidth]{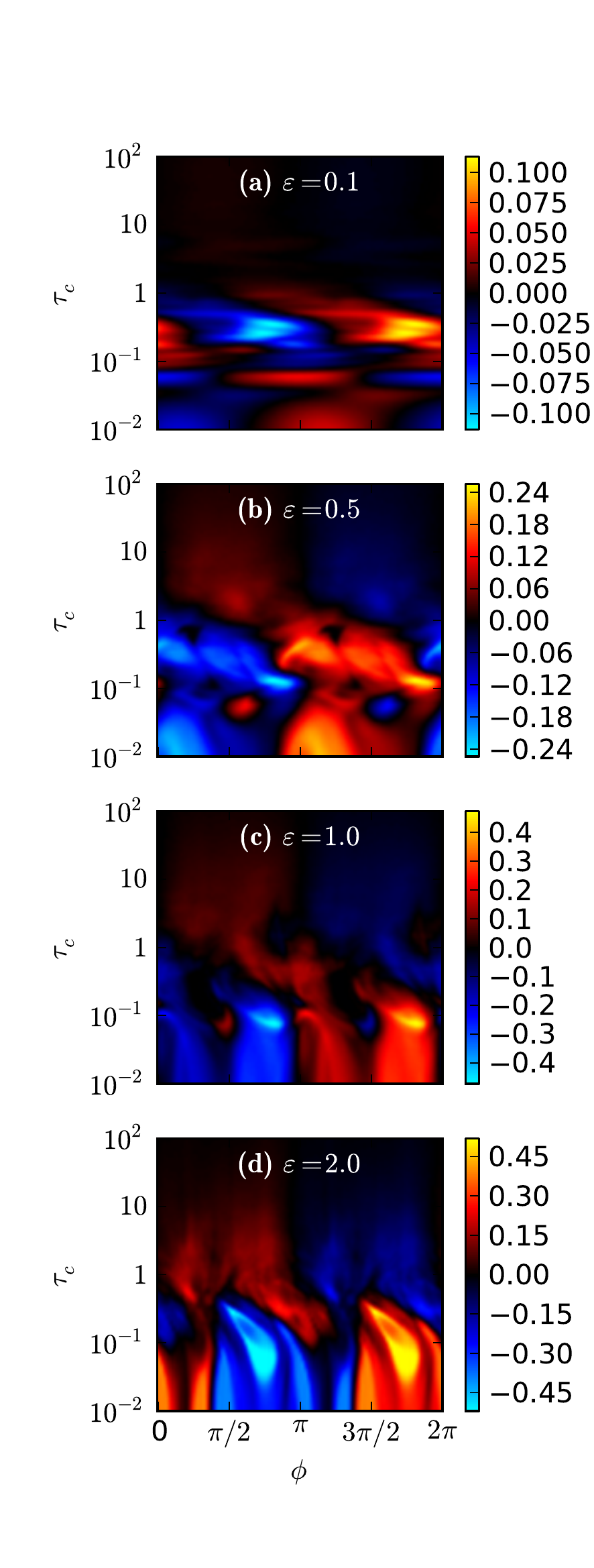} 
\caption{(color online) Influence of the second harmonic of the external force $G(t)$
on transport properties of the system (\ref{Gen}) is presented.
Dependence of the drift velocity on both the relative phase $\phi$ (horizontal axis) 
and the correlation time $\tau_c$ of thermal fluctuations (vertical axis) is depicted 
for increasing values of the amplitude $\varepsilon$ of the biharmonic 
driving $G(t)$  (top to bottom). 
The other rescaled parameters read
$a = 4.2$, $\omega = 4.9$, $\gamma = 0.9$, $d_0 = 0.001$.
}
\label{fig2}
\end{figure}

\section{Impact of correlated thermal noise on transport}

In the  long time limit,  the  averaged velocity can be presented in the form of a series of all possible harmonics, namely,  
\begin{eqnarray}
\label{asym}
\lim_{t\to\infty} \langle {\dot X} \rangle =  v + v_{\omega}(t) + v_{2\omega}(t) + \dots, 
\end{eqnarray}
where $v$ is a  dc (time-independent) component and $ v_{n \omega}(t)$ are time-periodic functions which time average over a basic period are  zero.  In figures we present only  the dc component  $v$ of the long time,  averaged velocity.  From the symmetry property of  Eq.  (\ref{Gen}) it follows when $v$ vanishes, for details see \cite{Denisov2010,sym,niurka,renzoni,mach2010}.  
In particular, in Ref. \cite{niurka} it has been shown that if the  force $F(t)$ in Eq. (\ref{F}) has the form 
\begin{equation}
\label{f}
F(t) =  \epsilon_1  \cos(q\omega t + \phi_1)  + \epsilon_2  \cos(p\omega t + \phi_2) 
\end{equation}
and $(p, q)$ are two coprime integers such that $p+q$  is odd,  the asymptotic velocity 
can be approximated   by the expression 
\begin{equation}
\label{vas}
v=  B \epsilon_1^p   \epsilon_2^q \cos(p \phi_1 - q \phi_1 + \theta_0)
\end{equation}
provided the  amplitudes $ \epsilon_1$ and $ \epsilon_2$ are sufficiently small. The quantities  $B$ and $\theta_0$ depend on the parameters of the model and $\omega$ but neither on the amplitudes nor on the phases. 

We remind that in the Hamiltonian regime  Eq. (\ref{Gen}) reduces to the form 
\begin{eqnarray}
\label{Ham}
m\ddot x(t)  = - U'(x(t), t) 
\end{eqnarray}
and  for small amplitudes the dc velocity takes the sine-like  form \cite{Breymayer1984,flachEPL}
\begin{eqnarray}
\label{vHam}
v \propto \sin \phi. 
\end{eqnarray}
Mechanism of the transport generation in Hamiltonian systems is explained in Ref. \cite{denis}. 
The Hamiltonian regime can be realized  in two case: (i)  in the dissipationless regime when  $\gamma=0$ and $D=0$, (ii) in the limit of long correlation time, $\tau_c >>1$.

For white noise case and in the weak damping  regime, the dissipation-induced phase lag $\phi_0$ occurs, i.e.    \cite{Breymayer1984,flachEPL}
\begin{eqnarray}
\label{weak}
v \propto \sin (\phi - \phi_0),  
\end{eqnarray}
where the phase lag $\phi_0$ is determined by dissipation and vanishes in the Hamiltonian limit.

We first   address the issue of whether, and to which extent,
the non-zero correlation time $\tau_c$ of thermal fluctuations can  influence transport properties. 
We thus start our analysis by studying the asymptotic dc velocity  $v$ as
a function of both the time-symmetry breaking  phase $\phi$   and the correlation time  $\tau_c$. 
 Our results are shown in Fig.  2 for selected values of the driving amplitude $\varepsilon$ 
of the second harmonics. A general conclusion from all cases presented in this figure is the 
destructive influence of the strongly  correlated noise: for long correlation time  $\tau_c$ the dc velocity is much smaller (virtually zero) than for the small-to-moderate correlation time. 
   Let us note that  the limit of  the long correlation time ($\tau_c >> 1$), corresponds to the 
Hamiltonian regime in which  transport is non-effective in comparison with the dissipative regime. 
A more accurate  inspection reveals regions of   weakly correlated noise  where  reach diversity of transport characteristics can be observed.   To identify them we search the section $\phi = const.$ of a constant phase  in Fig. 2.  
The results are depicted in Fig. 3 where the average velocity is plotted as a function of the correlation time for three  selected 
values of the amplitude $\varepsilon = 0.1, 1, 2$ (the same as in panels a, c and d  of  Fig.  \ref{fig2}). The phase is fixed at the values  
$\phi = 1.65$ and $\phi=2.41$. In  Fig. 3, one  can identify the so termed  current reversal phenomenon: the velocity 
changes its sign as one of the parameters is varied. Here,  even the multiple current reversal  can be observed by changing  the noise 
correlation time.  E.g.  for the amplitudes $\varepsilon = 1 $ in Fig. 3(a), the  average velocity is negative for white noise, i.e. when $\tau_c =0$. If the correlation time $\tau_c$ increases, the velocity approaches zero value: the particle  does 
not exhibit directed transport.  Upon further  increasing   of $\tau_c$, the velocity starts to increase  to a positive-valued  local 
maximum. Next, it  again starts to decrease reaching  a negative-valued  local minimum. For further increase of  $\tau_c$, the velocity 
tends again to a positive-valued local maximum. Finally, with increasing $\tau_c$ it monotonically decreases  towards  smaller and smaller positive values.   
\begin{figure}[tbp]
\includegraphics[width=0.99\linewidth]{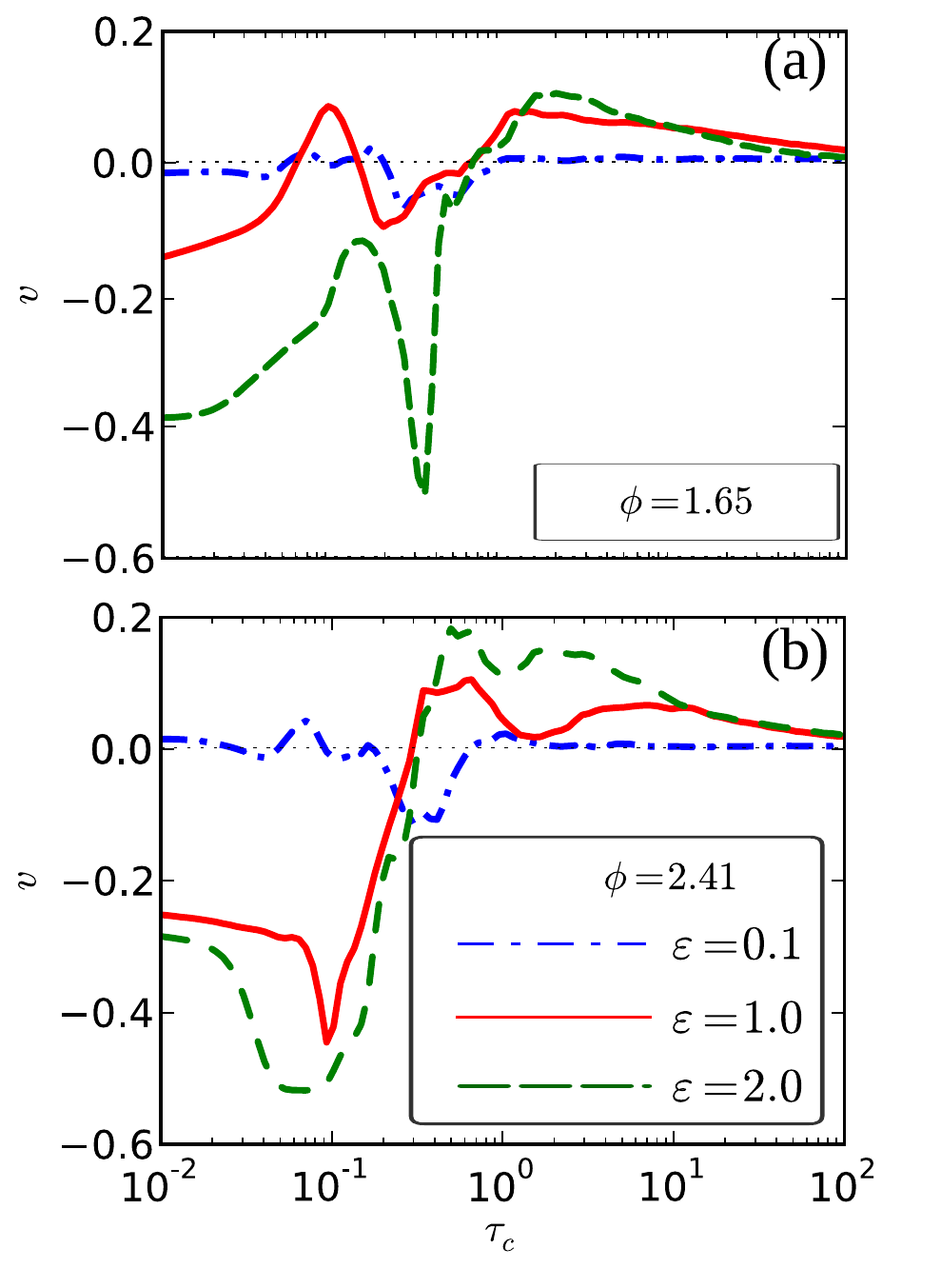} 
\caption{(color online) The stationary averaged velocity {\it vs} the correlation time 
$\tau_c$  of thermal noise for three values of the relative amplitude $\varepsilon$ 
of the second harmonics and two values of the relative phase $\phi$.  Other parameters are:  
$a = 4.2$, $\omega = 4.9$, $\gamma = 0.9$, $d_0 = 0.001$.
}
\label{fig3}
\end{figure}
\begin{figure}[tbp]
\includegraphics[width=0.99\linewidth]{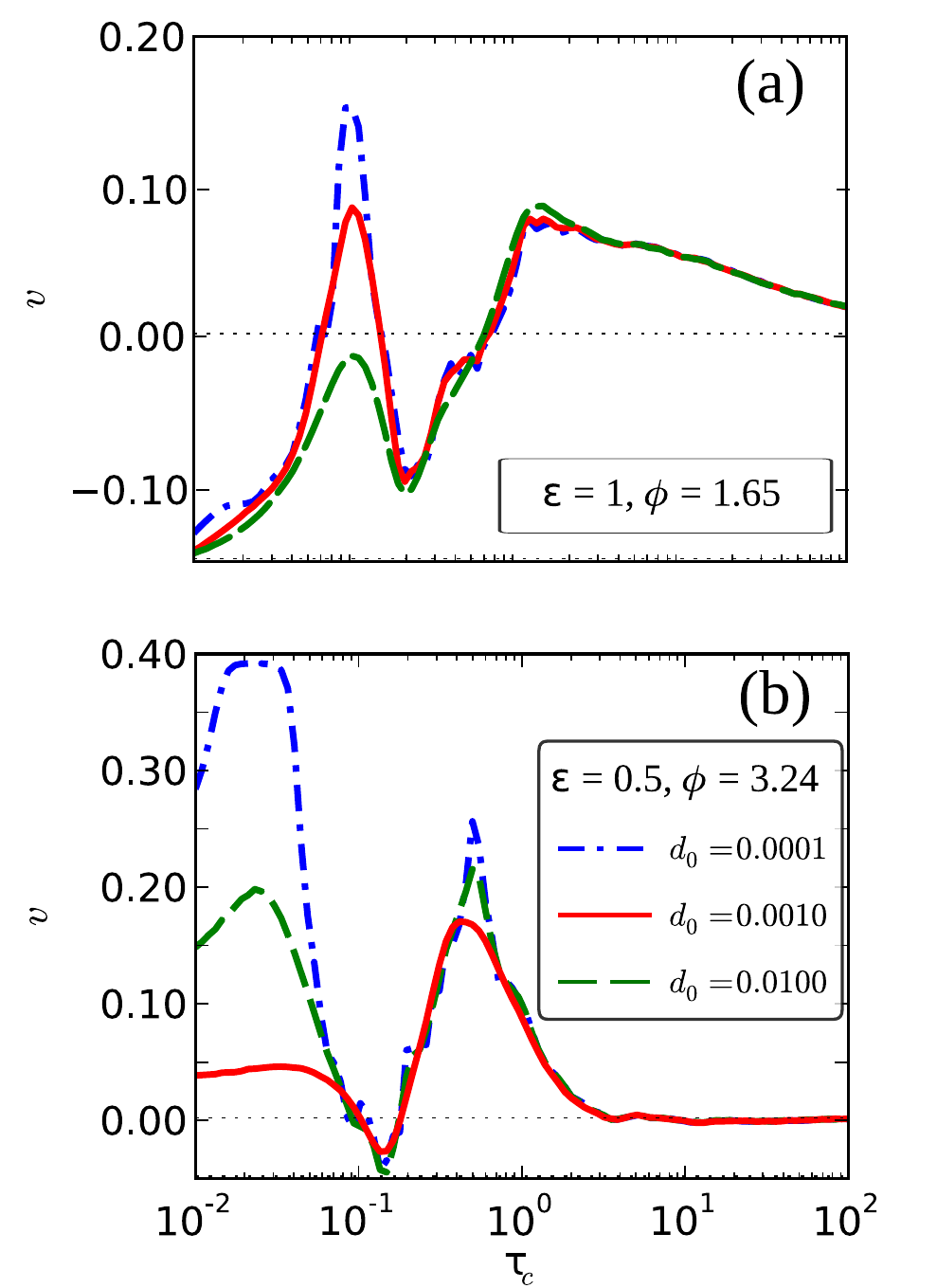} 
\caption{(color online) The stationary averaged velocity {\it vs} the correlation time 
$\tau_c$  of thermal noise for three values of the relative temperature $d_0 = 0.0001,0.001,0.01$ 
and two sets of pairs $\varepsilon = 1, \phi = 1.65$ (panel a) and $\varepsilon = 0.5, \phi = 3.24$ (panel b).  
Please note the destructive role of the small correlation time $ \tau_c $ on panel (a) and the antagonistic
effect of the constructive impact of $ \tau_c $ in panel (b). For large values of $ \tau_c $ the average velocity is always much smaller and virtually  is zero. Other  parameters are:  
$a = 4.2$, $ \varepsilon = 1 $, $\omega = 4.9$, $\gamma = 0.9$.
}
\label{fig4}
\end{figure}
As the next  point of analysis, we ask about the positive role of  weakly correlated noise, 
i. e. whether small $\tau_c$ can intensify transport by enhancing the absolute value of the 
average velocity.  
The answer is yes. As an  example, consider the case $\varepsilon = 2$ in Fig. 3(b). 
For $\tau_c = 0$, the velocity is negative  and its absolute value is relatively large. If $\tau_c$ starts to increase,
the absolute value of the velocity also increases attaining  a global maximum at some fixed 
value of $\tau_c$ (of order $10^{-1}$). In this example, the weakly correlated  noise  enhances transport in some interval 
of $\tau_c$. On contrary, for the same amplitude $\varepsilon =2$ 
but different value of the phase $\phi = 1.65$, if $\tau_c$ starts to increase from zero, 
the absolute value of the velocity decreases, see the case  $\varepsilon =2$  in Fig. 3(a).

The influence of temperature is depicted in Fig. \ref{fig4}, where two various regimes are illustrated.  In panel (a) we show the same curve as in panel (a) of Fig. 3 (the case $\varepsilon = 1$ and $d_0=0.001$).  In this regime  we can note  the most pronounced  influence  of temperature  
in vicinity of the first maximum: if temperature increases, the first maximum 
is significantly reduced and  velocity even changes its sign.  So, by  changing the temperature one can obtain the current reversal, see the curves for $d_0=0.001$ and $d_0=0.01$ in panel (a). 
We have searched 
a wide part of the parameter space  in order to locate regions of temperature robustness. The conclusion is that for large values 
of the correlation time (in most cases $\tau_c > 1$) the system is resistant to increase of temperature, at least  to values  presented 
in Fig. \ref{fig4}. On the contrary, for small correlation times,  there are regimes where the system is sensitive to changes of temperature and increase of 
temperature responses in smaller values of the absolute value of velocity, see panel (b) of Fig.  \ref{fig4}. It is a region of destructive role of temperature but the constructive role of short correlation times of thermal fluctuations: a first  increase of $\tau_c$ from zero  results in increase of the velocity.    
For moderate and long correlation time, the temperature smooths the dependence on the correlation time similarly as in both panels of  Fig. \ref{fig4}.


\begin{figure}[t]
\includegraphics[width=0.9\linewidth]{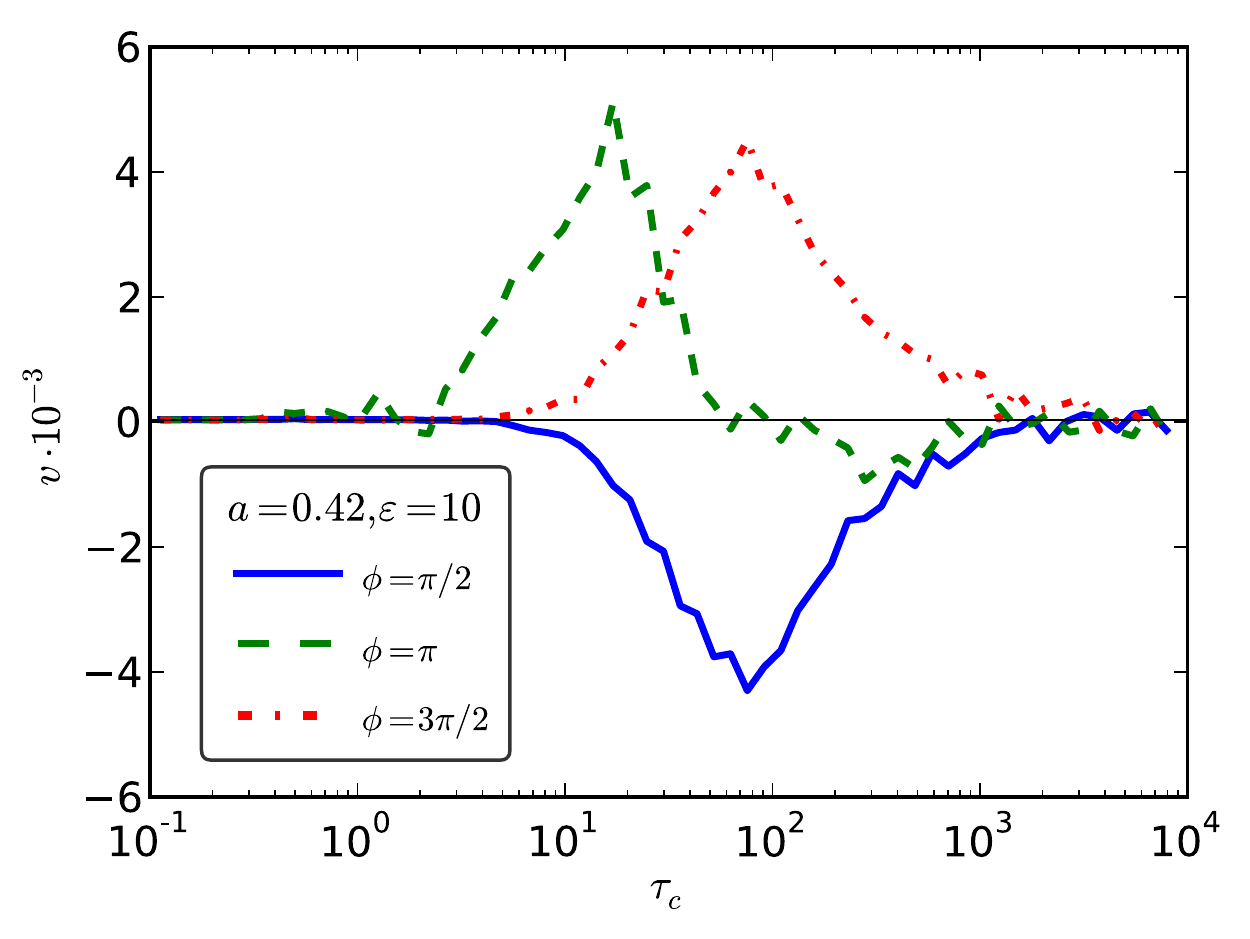} 
\caption{(color online) The stationary averaged velocity vs the correlation time $ \tau_c $ of thermal 
noise for three values of the relative phase $ \phi $ = $ \pi/2 $ blue (solid) line, $ \pi $ green 
(dashed) line, $ 3\pi/2 $ red (dashed--dotted) line.
Other parameters are: 
$a$ = 0.4, $ \varepsilon = 10 $, $ \omega $ = 4.9, $ \gamma $ = 0.9, $ d_0 $ = 0.001.
}
\label{fig5}
\end{figure}

Finally, in  Fig.  \ref{fig5} we present a regime where the amplitude of the second harmonics is much greater than the first harmonics (here 10 times greater).  In such a case, the degree of  symmetry breaking is very small: the 
influence of the first harmonics is much smaller than the second one (remember that  if any of two harmonics is absent, there is no transport in the system).  Therefore  the dc velocity is expected to be extremely small (practically, it is zero).  It is true but only for short correlation times (from  0 to 1 in Fig. 5).  When we, however, increase $ \tau_c $
a little bit more and pass the value of about 1, the significant net transport occurs. For some values of the phase $\phi$  the velocity is  negative velocity ($ \phi = \pi /2 $); for other values of $\phi$ it  shows
the positive response ($ \phi = 3\pi /2$). In some other cases it  exhibits the velocity  reversal phenomenon as e.g. for $ \phi = \pi $.

\begin{figure}[tbp]
\includegraphics[width=0.99\linewidth]{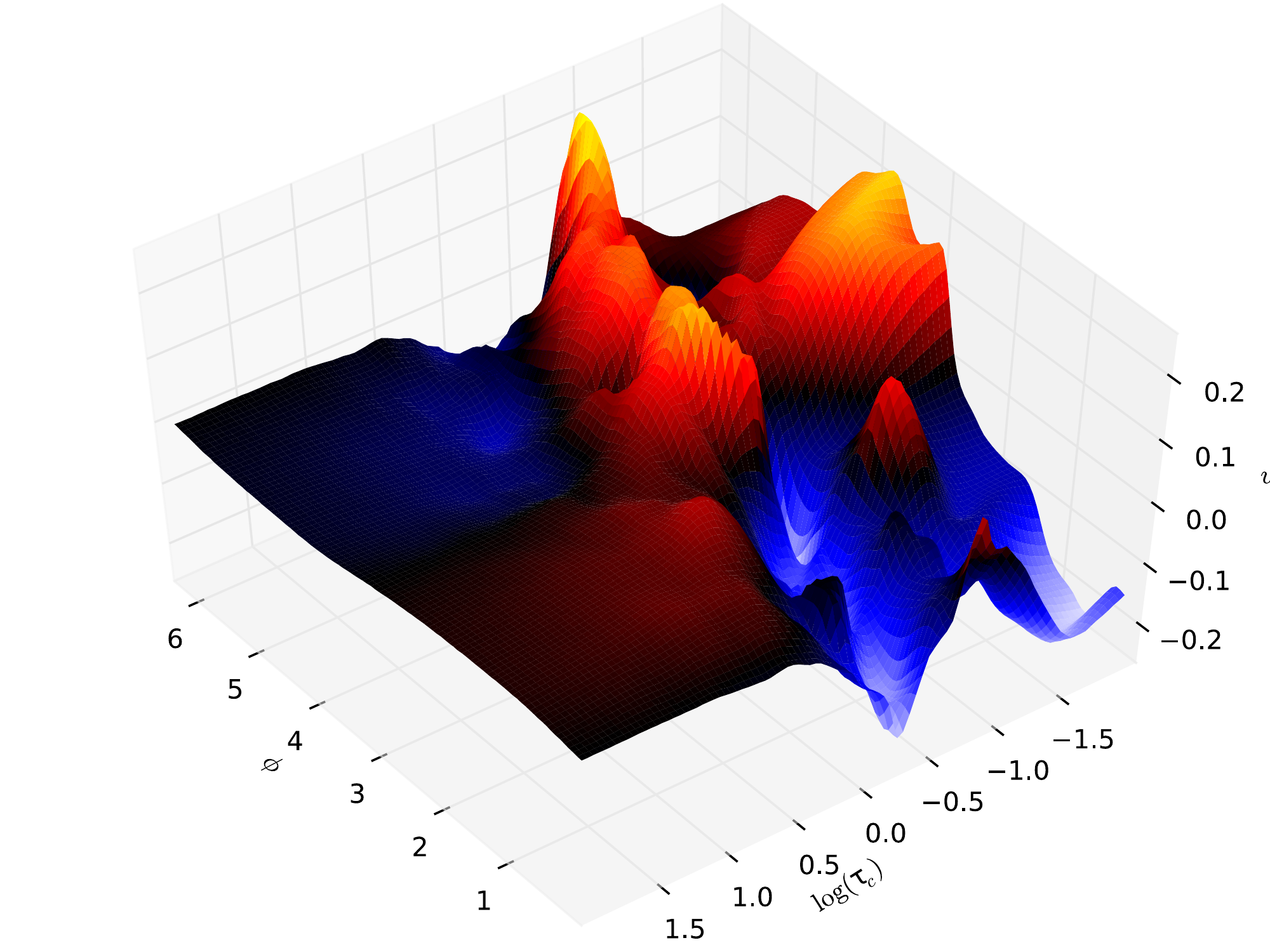} 
\caption{(color online) Illustration of role of the thermal noise correlations and the relative phase of driving two harmonics on transport of the Brownian particle. It is a regime where long correlations of thermal noise destruct  transport. The parameters: 
$a = 4.2$, $\varepsilon = 0.5$, $\omega = 4.9$, $\gamma = 0.9$, $d_0 = 0.001$
}
\label{figc1}
\end{figure}

We have mainly concentrated the analysis  on impact of the correlation time of thermal noise on  transport properties of the Brownian particle.  Other parameters modify directed movement as well.  
We do not present all varieties but yet  we refer  to our webpage  \cite{web}  where  the  more detailed analysis  is presented.  
\section{Summary}

The richness and diversity of influence of non-zero correlation time of thermal noise on 
non-Markovian dynamics determined by Eq. (\ref{Gen})  is summarized  and visualized in 
Fig. 6.  This  picture  looks like a landscape 
after intensive eruption of volcanoes: there are mountains with sharp summits and 
volcanoes with deep craters. 
This landscape  can  continuously be deformed to other complicated  landscapes  by changing 
parameters of the model. If you are long-suffering, you can  discover your own original landscapes 
of non-Markovian dynamics driven by biharmonic signals. The diversity of  
subtle structures hidden in Eq. (\ref{Gen}) is infinitely large.  It comes to our mind a loose analogy to 
a Mandelbrot set which  conceals  beautiful forms of Julia sets, fractals and  extremely difficult mathematics.   

\begin{acknowledgements}
The authors thank M. Januszewski for preparing the precise program  
(http://gitorious.org/sdepy) that we have used for numerical calculations.
The work supported  by  the ESF  Program "Exploring the Physics of Small Devices". 
\end{acknowledgements}


\begin{thebibliography}{99} 

\bibitem{RMP2009} P. H\"anggi and F. Marchesoni, Rev. Mod. Phys. {\bf 81}, 387 (2009).

\bibitem{schneider}      C.  E. Skov and E.  Pearlstein,   Rev. Sci. Inst. {\bf 35}, 962 (1964);  
W.  Schnerder and K. Seeger, Appl. Phys. Lett. {\bf 8}, 133 (1966). 

\bibitem{Borromeo2005a} 
F. Marchesoni, Phys. Lett. A {\bf 119}, 221 (1986); 
M. Borromeo and  F. Marchesoni, Europhys. Lett. {\bf 72}, 362 (2005); 
M. Borromeo, P. H\"anggi and  F. Marchesoni, J. Phys.: Condens. Matter {\bf 17}, S3709 (2005);
M. Borromeo and  F. Marchesoni, Phys. Rev. E {\bf 73},  016142 (2006) .

\bibitem{Breymayer1984}
H. J. Breymayer, Appl. Phys. A  {\bf 33}, 1 (1984).

\bibitem{igor}
Goychuk, I., H\"anggi, P.,  Europhys. Lett. {\bf 43}, 503 (1998). 

\bibitem{Renzoni2005} 
M. Schiavoni, L. Sanchez-Palencia,  F.  Renzoni,  and G.  Grynberg, 
   Phys. Rev. Lett. {\bf 90}, 094101 (2003); 
R. Gommers, S.  Bergamini,  F. Renzoni,    Phys. Rev. Lett. {\bf 95},
  073003 (2005); 
F. Renzoni, Cont. Phys. {\bf 46},  161 (2005).

\bibitem{Renzoni2008}
M. Brown and  F. Renzoni, Phys. Rev. A {\bf 77},   033405 (2008) .  

\bibitem{Denisov2010}
S. Flach, O.  Yevtushenko,  and  Y.  Zolotaryuk, Phys. Rev. Lett. {\bf 84}, 2358 (2000); 
S. Denisov, S. Flach and  P. H\"anggi, 
in: \emph{Nonlinearities in Periodic Structures and Metamaterials}, 
C. Denz, S. Flach, and Y. Kivshar, eds. Springer Series in Optical Sciences vol. 150 (2010) 181 Springer.

\bibitem{Monaco1990}
R. Monaco, J. Appl. Phys. {\bf 68}, 679 (1990). 

\bibitem{euro2}  P. H\"anggi, R. Bartussek, P. Talkner, and J. {\L}uczka, 
Europhys. Lett.  {\bf 35}, 315 (1996). 

\bibitem{kubo66}
R. Kubo, Rep. Prog. Phys. {\bf 29}, 255 (1966).

\bibitem{stewart}
W.C. Stewart, Appl. Phys. Lett. {\bf 12},  277(1968); 
D. E. McCumber, J. Appl. Phys.   {\bf  39}, 3113 (1968); 
R.~L. Kautz, Rep. Prog. Phys. {\bf 59},  935 (1996).  

\bibitem{junction} A. Barone and  G. Patern\`o, {\it Physics and
    Application of the Josephson Effect}, Wiley, New York, (1982).

\bibitem{jungACP}
P. H\"anggi and  P. Jung, Adv. Chem. Phys. {\bf 89}, 239 (1995).

\bibitem{chaos}  J. {\L}uczka, Chaos {\bf 15}, 026107 (2005).

\bibitem{goychuk2005}
I. Goychuk and P. H\"anggi, Adv. Phys. {\bf 54}, 525 (2005).

\bibitem{zwan} R. Zwanzig, J. Stat. Phys. {\bf 9}, 215 (1973).


\bibitem{LNP}
P. H\"anggi, Lect. Notes Phys. {\bf 484}, 15 (1997).

\bibitem{ornst} G.E. Uhlenbeck and L.S. Ornstein, Phys. Rev. {\bf 36}, 823 (1930). 

\bibitem{lutz} L. Schimansky-Geier, C. Z\"ulicke,
Z. Phys. B {\bf 79}, 451 (1990). 


\bibitem{srokowski} T. Srokowski and M. Płoszajczak, Phys. Rev. E {\bf 57}, 3829 (1998) 

\bibitem{goj} I.  Goychuk, Phys. Rev. E {\bf 80}, 046125 (2009) 

\bibitem{kupfer} R. Kupferman, J. Stat. Phys. {\bf 114}, 291 (2004).

\bibitem{bao} J. -D. Bao,  Phys. Rev. E {\bf 69}, 016124 (2004).

\bibitem{pha} J. {\L}uczka, Physica A {\bf 153}, 619 (1988). 

\bibitem{KosLucHan2009} M. Kostur, J. \L uczka, and P. H\"anggi, Phys. Rev. E {\bf 80}, 051121  (2009). 

\bibitem{bio}
L. Machura, M. Kostur and  J. {\L}uczka, Biosystems {\bf 94}, 253 (2008). 

\bibitem{PhysRevA.45.600}  
R. L. Honeycutt, Phys. Rev. A {\bf 45} 600 (1992).

\bibitem{cuda}
M. Januszewski and  M. Kostur, Comput. Phys. Commun. {\bf 181} 183 (2010).

\bibitem{sym} E. Neumann and A. Pikovsky, Eur. Phys. J. B {\bf 26}, 219 (2002); 
R. Chac\'on  and  N. R. Quintero, BioSystems {\bf 88},  308 (2007). 
\bibitem{niurka} 
N.  R. Quintero, J.  A. Cuesta, and R.  Alvarez-Nodarse, Phys. Rev. E  {\bf 81}, 030102 R (2010). 

\bibitem{renzoni} R. Gommers, S. Bergamini, and F. Renzoni, Phys. Rev. Lett. {\bf 95},
073003 (2005); A. B. Kolton and F. Renzoni, Phys. Rev.  A {\bf 81}, 013416 (2010) and refs therein.

\bibitem{mach2010} L. Machura, M. Kostur and  J. {\L}uczka, Chem. Phys. (2010), 
 doi: 10.1016/j.chemphys.2010.03.008. 

\bibitem{flachEPL}  
O. Yevtuschenko,  S. Flach, Y. Zolotaryuk and A. A. Ovchinnikov, Europhys. Lett. 54 (2001) 141. 
\bibitem{denis} S. Denisov and S. Flach, Phys. Rev. E  {\bf 64}, 056236 (2001).  
\bibitem{web} http://fizyka.us.edu.pl/biharmonic


\end{thebibliography}
\end{document}